\documentclass[prc,twocolumn,nofootinbib]{revtex4-2}

\usepackage{color}
\usepackage{graphicx}
\usepackage{dcolumn}
\usepackage{bm}
\usepackage{braket}
\usepackage{amsmath}
\usepackage{here}

\usepackage[normalem]{ulem}  

\renewcommand\sout{\bgroup\color[rgb]{1,0.75,0.8} \ULdepth=-.5ex \ULset}

\renewcommand{\vec}[1]{\mbox{\boldmath $#1$}}

\begin{document}


\title{Microscopic description of cluster decays based on the generator coordinate method}

\author{K. Uzawa, K. Hagino, and K. Yoshida}

\affiliation{%
 Department of Physics, Kyoto University, Kyoto 606-8502, Japan
}%

\date{Received\today}

\begin{abstract}

\noindent\textbf{Background:}
While many phenomenological models for nuclear fission have been developed,
a microscopic understanding of fission has remained one of
the most challenging problems in nuclear physics.

\noindent\textbf{Purpose:}
We investigate an applicability of the generator coordinate method (GCM)
as a microscopic theory
for cluster radioactivities of heavy nuclei, which can be regarded
as a fission with large mass asymmetry, that is, a
phenomenon in between fission and $\alpha$-decays.

\noindent\textbf{Methods:}
Based on the Gamow theory,
we evaluate the preformation probability of a
cluster with GCM while
the penetrability of the Coulomb barrier is estimated with a potential model.
To this end, we employ Skyrme interactions and solve the one-dimensional
Hill-Wheeler equation with the mass octupole field.
We also take into account
the dynamical effects of the pairing correlation
using BCS wavefunctions constructed
with an increased strength of the pairing interaction.

\noindent\textbf{Results:}
We apply this scheme to
the cluster decay of $^{222}$Ra, i.e., $^{222}$Ra$\to^{14}$C+$^{208}$Pb,
to show that the experimental decay rate can be reproduced
within about two order of magnitude.
We also briefly discuss the cluster radioactivities of
the $^{228}$Th and $^{232}$U nuclei.
For these actinide nuclei, we find that the present calculations reproduce
the decay rates with the same order of magnitude and within two or three order of magnitude, respectively.

\noindent\textbf{Conclusions:}
The method presented in this paper
provides a promising way
to describe microscopically cluster decays of heavy nuclei.

\end{abstract}

\maketitle

\section{INTRODUCTION}

Nuclear fission is an important phenomenon in various areas of physics,
including productions of neutron-rich nuclei,
the r-process nucleosynthesis,
and syntheses of superheavy elements
\cite{vandenbosch,krappe,frobrich,schunck2016}.
While many phenomenological models for fission have been developed,
a microscopic understanding of fission has remained one of
the most challenging problems in nuclear physics \cite{bender2020}.
In the fission process, many degrees of freedom are involved
during a shape evolution of a fissioning nucleus.
Since it is difficult
to incorporate all the degrees of freedom,
it is essential to extract
appropriate degrees of freedom for fission.
Nuclear deformation parameters are often used for this purpose.

In addition to nuclear deformation, the nuclear superfluidity \cite{superfluidity}
also plays an important role in describing the fission process, see e.g.
Refs. \cite{schutte1975,schutte1978,scamps2015,washiyama2021}.
In particular, the role of dynamical pairing \cite{Moretto74}
has attracted lots of attention in recent years
\cite{giuliani2014,sadhukhan2014,guzman2018}.
To explicitly take into account the pairing
dynamics,
a pair hopping model was proposed
in Ref. \cite{barranco1990}.
In this model,
a nucleus goes into a shape evolution by hopping from one
Hartree-Fock configuration to a neighboring configuration
by a residual pairing interaction.
This model has been successfully applied to recent alpha decay
experiments for high spin isomers \cite{rossamem2014,clark2018,clark2019}.
Based on a similar idea to the pair hopping model,
a more microscopic approach based on a many-body Hamiltonian was
investigated in Refs. \cite{bertsch2020,hagino2020a}.
Advantages of this approach are
i) it is easy to connect to reaction theories \cite{bertsch2020}
and ii) a collective inertia for fission does not need to be evaluated explicitly.
Within this model, the effect of the dynamical pairing was
considered by introducing
the maximum coupling approximation \cite{hagino2020b},
in which the basis states are constructed by increasing the pair
correlation.

In this connection, an interesting phenomenon to explore is a cluster
radioactivity, such as an emission of $^{14}$C from a heavy nucleus.
This phenomenon can be regarded as a phenomenon
in between spontaneous
fission and $\alpha$ decays.
This is a unique phenomenon, in which
many-body effects
are much more important than $\alpha$ decays,
while the matching of a many-body wave function to an external region
is much simpler than that for spontaneous fission.
The cluster radioactivity was observed for the first time
in 1984 in the decay of $^{223}$Ra emitting the $^{14}$C cluster \cite{rose1984}.
Since then, several cluster emission decays
have been observed by now \cite{textbook},
in which
a daughter nucleus tends to be
$^{208}$Pb or its neighbors.
Notice that the cluster radioactivities may be regarded as a spontaneous fission
with large mass asymmetry.
See Refs. \cite{warda2011,warda2018,matheson2019}
for recent studies along this line on the cluster radioactivities based on the density
functional theory.
Even though the branching ratio of the cluster decays
to alpha decays is usually considerably small, it has been pointed out that
the cluster decay may become a dominant decay mode of superheavy
nuclei \cite{warda2018,matheson2019,poenaru2012}.

In this paper,
we apply a similar approach to Refs. \cite{bertsch2020,hagino2020a,hagino2020b}
to
cluster radioactivities of heavy
nuclei.
While
Refs. \cite{bertsch2020,hagino2020a,hagino2020b} used a
schematic many-body Hamiltonian,
we here employ
a realistic energy functional of Skyrme type.
To this end,
we take into account the non-orthogonality
of many-body configurations at different shapes by the generator coordinate
method (GCM).
Also, we consider couplings
among all the configurations in a model space, not restricting to the nearest
neighbor couplings.

The paper is organized as follows.
In Sec. I\hspace{-1pt}I,
we detail the theoretical method for the cluster radioactivities
based on the generator coordinate method.
In Sec. I\hspace{-1pt}I\hspace{-1pt}I, we present results for the
cluster decay of $^{222}$Ra, $^{228}$Th and $^{232}$U as typical examples.
We compare the calculated decay rates with the experimental data as well as
with other theoretical calculations. We also discuss the role of dynamical
pairing in cluster decays. We then summarize the paper in
Sec. I\hspace{-1pt}V.

\section{Cluster decays based on GCM}

For a theoretical description of cluster decays,
two types of approaches have been employed\cite{textbook}, either based on the Gamow
theory for $\alpha$ decays \cite{gamow1928} or
on models for spontaneous fission.
In the former, it is assumed that a cluster is preformed
in a mother nucleus and then it tunnels through the Coulomb
barrier \cite{blendowske1987,delion1994}.
In this theory,
a decay rate is expressed as
\begin{equation}
  w=S f P,
\label{gamow}
\end{equation}
where
$S$ is the preformation probability for a cluster to appear
in a mother nucleus,
$f$ is a barrier assault frequency, i.e., an attempt frequency,
and $P$ is the penetration probability of the Coulomb barrier.
A similar approach can be formulated also using the Fermi Golden
Rule \cite{barranco1990}.
On the other hand,
in the latter approach\cite{warda2011,warda2018,matheson2019},
a potential energy surface
and mass inertias for fission
characterized by nuclear deformation parameters
are calculated
based on theoretical models such as the liquid drop model
or the density functional theory.
The decay rate is estimated from the least action path
in the potential energy surface so obtained.

In this paper, we employ the Gamow theory to
compute the decay rates.
To this end,
we estimate the preformation probability $S$
based on the generator coordinate method (GCM), while
$f$ and $P$ based on a two-body potential model.
That is,
we carry out a microscopic calculation before the clusters are preformed
using the GCM, while we use a phenomenological two-body approach
after that.
In principle, we could use the microscopic density functional
theory also for the tunneling process. However, this would
require a large model space as well as a proper treatment
of the neck degree of
freedom \cite{warda2011,han2021} (see also Ref. \cite{lau2021}).
We thus leave it for a future study.

To calculate the preformation probability of a cluster, we first
solve the Hartree-Fock (HF) equation
with constraints on mass multipole moments.
The pairing correlation is also taken into account in the BCS approximation.
Following Ref. \cite{warda2011},
we use the mass octupole moment,
$Q_{3}=\sum_{i} r_i^3Y_{30}(\theta_i)$, for the constrained calculations.
For the particle-hole interaction,
we use the Skyrme interaction with the SkM*
\cite{bartel1982} and the SLy4 \cite{chabanat1998} parametrizations.
We solve the HF equation using the imaginary-time method with
the coordinate-space representation\cite{davies1980}.
We impose axial symmetry
and use the two-dimensional cylindrical mesh.

For the pairing interaction, we employ a volume-type
contact interaction,
\begin{equation}
V_{\rm pair}(\bm{r},\bm{r}')=V_{\tau}\,\frac{1-P_{\sigma}}{2}\delta(\bm{r}-\bm{r}'),~~(\tau=n,p)
\end{equation}
where $P_{\sigma}$ is the spin exchange operator.
The value of $V_{\tau}$ is determined to reproduce the empirical pairing gaps,
\begin{equation}
  \begin{split}
  &\Delta_n=-\frac{1}{2}[B(N-1,Z)+B(N+1,Z)-2B(N,Z)],\\
  &\Delta_p=-\frac{1}{2}[B(N,Z+1)+B(N,Z-1)-2B(N,Z)],\\
  \end{split}
\end{equation}
where $B(N,Z)$ is the measured binding energy \cite{ame2020} of the nucleus
with the neutron number $N$ and the proton number $Z$.
For a zero-range pairing interaction, the energy cutoff is
necessary to exclude high momentum components from the model space.
We use the smooth cut-off procedure
with a Fermi function \cite{bonche1985,krieger1990}.

Based on the idea of GCM\cite{ring},
we describe the decaying wave function as a superposition of the BCS
wave functions at different $Q_3$ values,
\begin{equation}
|\Psi\rangle
=\sum_i f(q_i)\hat{P}_Z\hat{P}_N
|\Phi(q_i)\rangle
\equiv\sum_i f(q_i)
|\Phi(N,Z,q_i)\rangle,
\label{eq:GCM}
\end{equation}
where
$|\Phi(q)\rangle$ is the BCS wave function at
$Q_3=q$, and
$\hat{P}_Z$ and $\hat{P}_N$ are the operators
to project the BCS wave function onto an eigenstate of the
proton and the neutron numbers, respectively.
The weight function $f(q_i)$ is determined by solving
the Hill-Wheeler equation,
\begin{equation}
  \begin{split}
&\sum_{j} \langle \Phi(N,Z,q_i) |H|\Phi(N,Z,q_j) \rangle f(q_j)\\
&=E\sum_{j} \langle \Phi(N,Z,q_i) |\Phi(N,Z,q_j) \rangle f(q_j).
\end{split}
\label{eq:HWeq}
\end{equation}
The preformation probability $S$ is then determined as
\begin{equation}
S=|g(Q_t)|^2,
\label{eq:S}
\end{equation}
where $Q_t$ corresponds to
the octupole moment at the
crossing point between the one-body and the two-body
configurations (see the discussion below Eq. (16)).
Here, the collective wave function $g(q_i)$ is defined as
\begin{equation}
  g(q_i)=\sum_{j} N^{1/2}(q_i,q_j)f(q_j),
\end{equation}
where $N(q_i,q_j)=\langle \Phi(N,Z,q_i)|\Phi(N,Z,q_j) \rangle$
is the overlap kernel and
$N^{1/2}(q_i,q_j)$
is the component of the matrix $N^{1/2}$.

In the usual GCM calculations, the basis function
$|\Phi(N,Z,q) \rangle$ is taken to be the local ground state at $q$.
Excitations during the decay process can also be taken into using
the configuration interaction (CI)
approach \cite{bertsch2020,hagino2020a,hagino2020b}.
That is, instead of Eq. (\ref{eq:GCM}), one can consider
\begin{equation}
|\Psi\rangle
=\sum_i\sum_k f_k(q_i)\hat{P}_Z\hat{P}_N
|\Phi_k(q_i)\rangle,
\end{equation}
where $\{|\Phi_k(q)\rangle\}$ is a set of
many-body wave functions at $q$, including
both the local ground state and excited states.
In Ref. \cite{hagino2020b}, an efficient way to include the excited
states has been proposed using the maximum coupling approximation.
In this approximation, one modifies the Hamiltonian
by increasing the strength of the pairing interaction by a factor $\alpha$,
\begin{equation}
H_{\rm mod}=H_{\rm HF}+\alpha H_{\rm pair},
\label{eq:Hmod}
\end{equation}
where $H_{\rm HF}$ and $H_{\rm pair}$ are the
particle-hole and the pairing parts of the Hamiltonian, respectively.
The local ground state of the modified Hamiltonian,
$|\Phi^{(\alpha)}(q)\rangle$, is then superposed as
\begin{equation}
|\Psi\rangle
=\sum_i f(q_i)\hat{P}_Z\hat{P}_N
|\Phi^{(\alpha)}(q_i)\rangle.
\end{equation}
The value of $\alpha$ can be determined so that the
decay rate is maximized.
Notice that
using the Thouless theorem\cite{thouless1960}
the wave function $|\Phi^{(\alpha)}(q_i)\rangle$ can be expressed as
\begin{equation}
  |\Phi^{(\alpha)}(q)\rangle
\propto
\prod_{i,j}\left(1+C^{(\alpha)}_{i,j}\alpha^{\dagger}_i\alpha^{\dagger}_j\right)|\Phi(q)\rangle,
\label{eq:thouless}
\end{equation}
where $\alpha_i^\dagger$ is a creation operator for quasi-particles,
thus it includes excited configurations in a specific way.

To compute the frequency $f$ and the penetrability $P$,
we consider a
phenomenological potential $V$ for the relative motion between the two
fragments,
\begin{equation}
V(r)=V_N(r)+V_C(r),
\label{pot}
\end{equation}
where $r$ is the relative coordinate, and $V_N$ and $V_C$ are
the nuclear and the Coulomb potentials, respectively.
For the Coulomb interaction, $V_C$,
we consider the potential for a uniformly charged sphere with the
radius $r_C$,
\begin{equation}
  V_C(r)=
  \begin{cases}
    \frac{Z_{1}Z_{2} e^2}{r} & \text{( $r>r_C$)} \\
    \frac{Z_{1}Z_{2} e^2}{r_C}(\frac{3}{2}-\frac{r^2}{2r^2_C}),                 & \text{( $r\leq r_C$)}
  \end{cases}
\end{equation}
where $Z_{1}$ and $Z_{2}$ are the proton number
of each fragment.
We take
$r_C=1.2(A_1^{1/3}+A_2^{1/3})$ fm
for the charge radius, where $A_1$ and $A_2$ are the
mass number of each fragment.
For the nuclear potential, $V_N$, we employ a Woods-Saxon potential
\begin{equation}
V_N(r)=-\frac{V_0}{1+\exp[(r-R_0)/a]},
\end{equation}
for which the radius parameter $R_0$ and the diffuseness parameter $a$
are taken from Ref. \cite{brogliawinther}.
The depth parameter $V_0$ is adjusted so that
the resonance energy of the potential $V$,
determined with the two-potential method \cite{gurvitz2004},
coincides with the
experimental $Q$-value \cite{ame2020}.
Even though there may be several uncertainties in the nuclear potential, especially
in the region
well inside the Coulomb barrier, we would expect that the order of magnitude
of a calculated decay rate is rather insensitive once the barrier
height is fixed. This is because the decay rate is largely determined
by the penetration probability of the Coulomb field;
even though the attempt frequency is sensitive to the nuclear potential,
it merely changes a multiplicative factor to the decay rate.

With the potential $V$ so determined, we
calculate $f$ and $P$ in the WKB approximation as \cite{aberg1997},
\begin{equation}
  \begin{split}
&f^{-1}=\frac{4\mu}{\hbar}\int^{r_1}_{r_0}\frac{dr}{k(r)}
\cos^2\left(\int^r_{r_0}k(r')dr'-\frac{\pi}{4}\right),\\
    &P=\exp\left(-2\int^{r_2}_{r_1}dr|k(r)|\right),
  \end{split}
\end{equation}
where $r_i(i=0,1,2)$ are the classical turning points, with $r_0$ and $r_2$
being the innermost and the outermost turning points, respectively.
$k(r)$ is the local wavenumber defined as $k(r)=\sqrt{2\mu[Q-V(r)]/\hbar^2}$,
where $\mu$ is the reduced mass for the relative motion between the clusters.
Notice that for proton radioactivities
the WKB approximation has been shown to agree well with
more quantal approaches such as
the Green function method and
the two-potential method
\cite{aberg1997}. We expect that the WKB approximation works even
better for cluster radioactivities with a larger reduced mass.

To compute the preformation probability $S$ according to
Eq. (\ref{eq:S}),
we convert the relative coordinate $r$ to the octupole moment $Q_3$ using an approximate formula given by \cite{warda2011} \footnote{This formula differs from Eqs. (9) and (10)
in Ref. \cite{warda2011} by a factor of $\sqrt{7/4\pi}$ due to the different
defintion for the octupole moment employed in this paper.}
\begin{equation}
Q_3(r)=\sqrt{\frac{7}{4\pi}}\frac{A_1A_2}{A_1+A_2}\frac{(A_1-A_2)}{A_1+A_2}r^3.
\label{Q3r}
\end{equation}
Notice that the Coulomb potential between the two clusters decreases
as a function of $r$, and thus $Q_3$, while
the one-body energy tends to increase.
Both curves will thus cross at
a certain octupole moment $Q_3$, which we label as $Q_t$.
In the region of $Q_3 > Q_t$, the total energy
becomes smaller when the mother nucleus splits into the two-body system.
We therefore regard that the cluster decays
happen via this configuration at $Q_3=Q_t$ and employ Eq.
(\ref{eq:S}) to estimate the cluster preformation probability.

\section{Results}

Let us now apply the model presented in the previous section
to the cluster decays of $^{222}$Ra, $^{228}$Th, and $^{232}$U
and numerically evaluate the decay rates.
To this end, we use the cylindrical mesh with
$r_i=(i-\frac{1}{2})\Delta r,~(i=1,2,...14)$
and $z_j=(j-\frac{1}{2})\Delta z,~(j=-13,-12,..26)$
with $\Delta r=\Delta z=0.8$ fm
for the Hartree-Fock+BCS calculations. In addition to the constraint of the mass octupole moment $Q_3$, we also impose a constraint on $\langle z \rangle=0$ in order to fix the
position of the center of mass.

\subsection{$^{222}\mathrm{Ra}$}

\begin{figure}
\includegraphics[width=8.6cm]{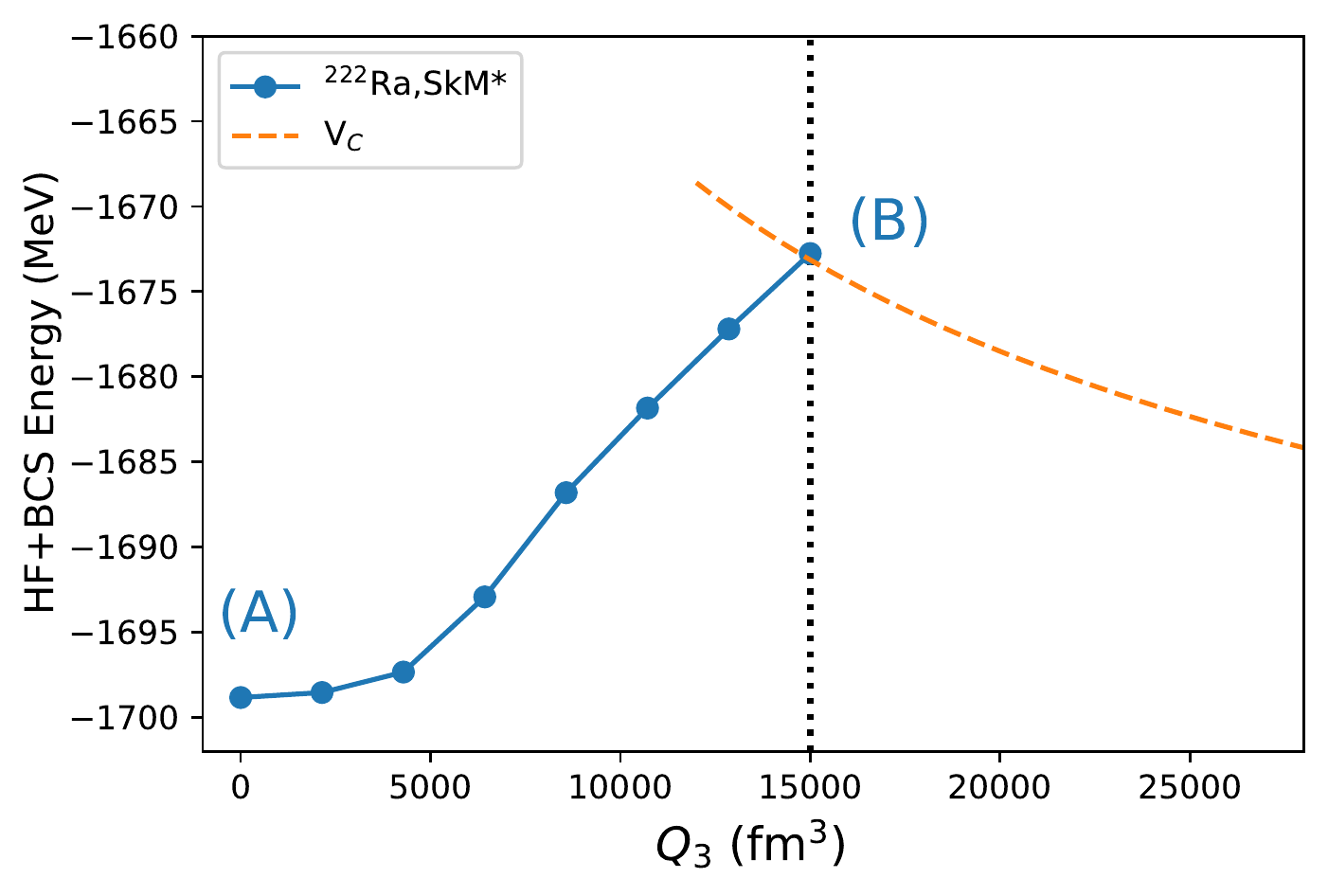}
\caption{
The Hartree-Fock (HF) + BCS energy obtained with the SkM$^*$ interaction
for the $^{222}\mathrm{Ra}$ nucleus as a function of
the mass octupole moment, $Q_3$.
The configuration (A) at $Q_3=0$ fm$^3$ corresponds to the
ground state in the HF+BCS approximation.
The dashed line represents
the Coulomb potential for the two-body system $^{14}$C+$^{208}$Pb.
This approaches
$E_{\rm gs}-Q$ at large $Q_3$, where $E_{\rm gs}$ is the energy for the
configuration (A).
The configuration (B)
at $Q_3$=15000 fm$^3$ corresponds to
the cluster configuration where the HF + BCS energy
crosses the dashed line (see the vertical dotted line).
}
\end{figure}

\begin{figure}
\includegraphics[width=8.6cm]{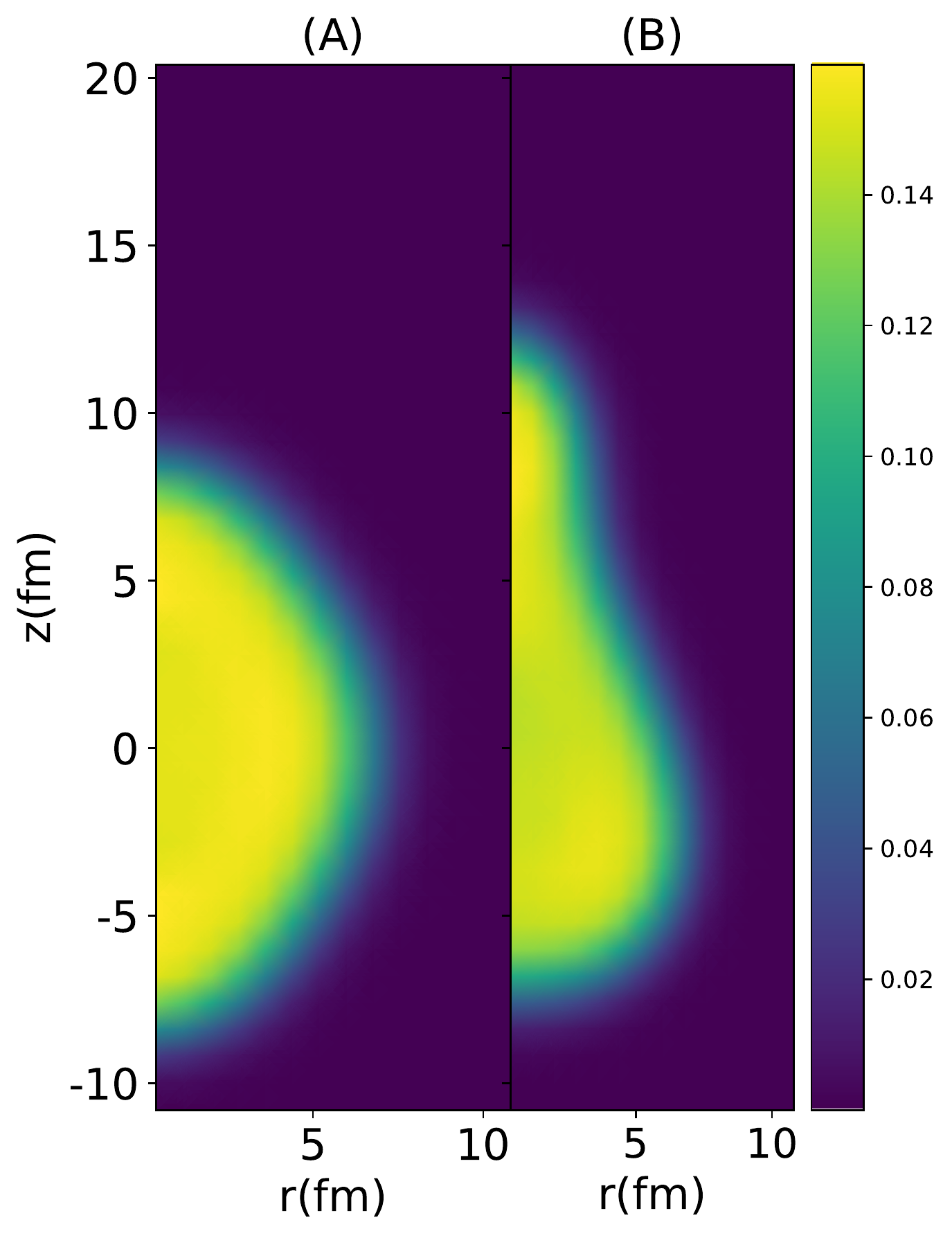}
\caption{The density distribution of the $^{222}\mathrm{Ra}$ nucleus
for the configurations (A) and (B) shown in Fig. 1.}
\end{figure}

\begin{figure}
\includegraphics[width=8.6cm]{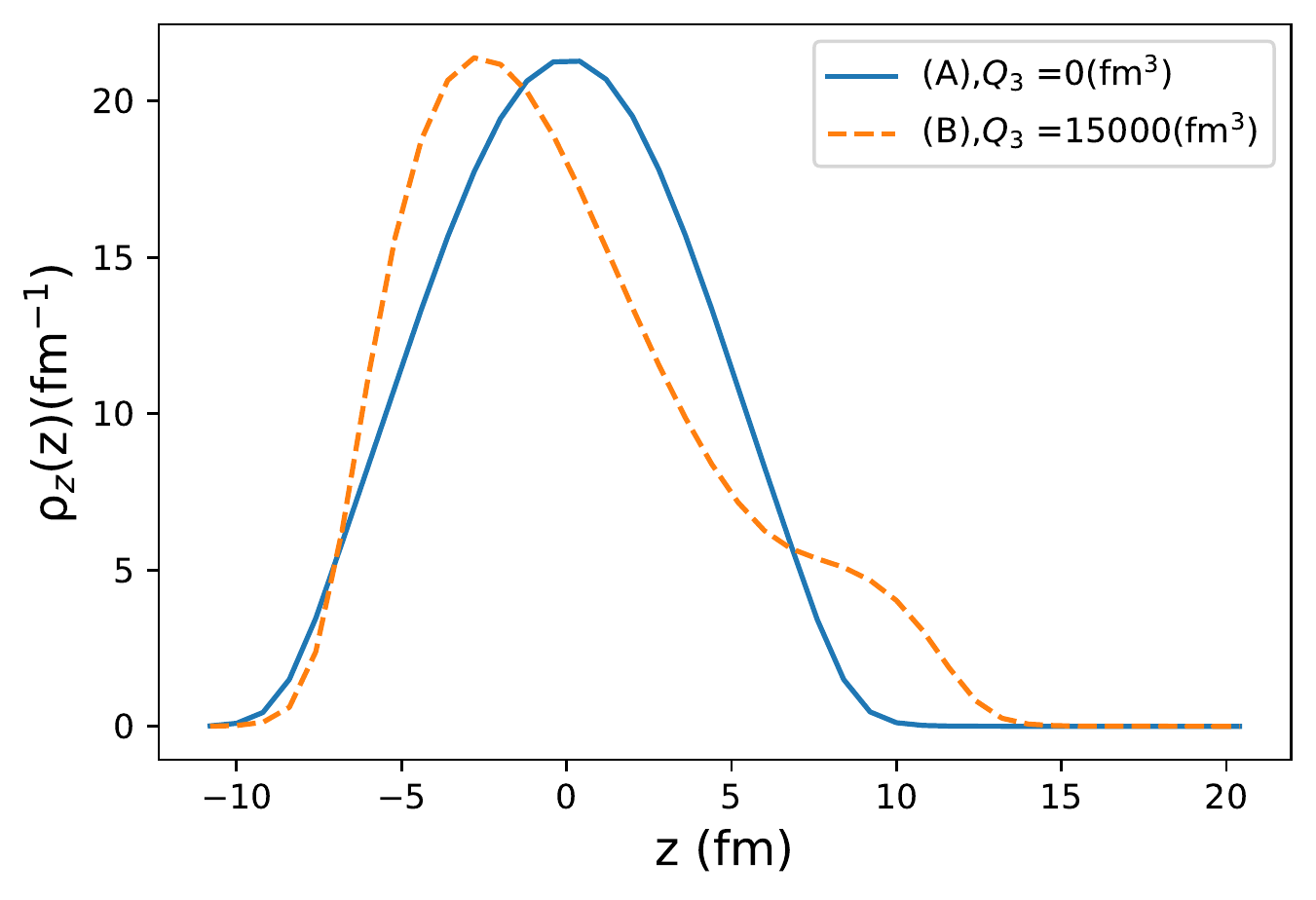}
\caption{
The density distributions
corresponding to those shown in Fig. 2
as a function of $z$.
Those are obtained by integrating the densities
in the $x$ and $y$ directions.
}
\end{figure}

We first discuss the decay of $^{222}$Ra $\to ^{14}$C+$^{208}$Pb,
whose $Q$-value is $Q=33.05$ MeV \cite{ame2020}.
We solve the HF + BCS equation with  $V_{p}=-398.0$ MeV fm$^{3}$ and $V_{n}=-280.0$ MeV fm$^{3}$ for the SkM$^*$ interaction, and $V_{p}=-420.0$ MeV fm$^{3}$ and $V_{n}=-320.0$
MeV fm$^{3}$ for the SLy4 interaction.
The blue solid line in
Fig. 1 shows the
HF+BCS energy of the $^{222}$Ra nucleus
obtained with the SkM$^*$ interaction
as a function of the mass octupole moment, $Q_3$.
The orange dashed line denotes the Coulomb potential,
$V_{C}(r(Q_3))$, shifted by $E_{\rm gs}-Q$,
where $E_{\rm gs}$
is the energy of the ground state in
the HF+BCS approximation.
In the ground state, denoted by (A) in the figure,
$^{222}$Ra is not octupole deformed.
As the octupole deformation is developed,
the energy increases
and eventually crosses the dashed line at the configuration (B).
The density distributions for the configurations (A) and (B)
are shown in Fig. 2.
See also Fig. 3 for the densities integrated in the $x$ and $y$ directions,
$\rho_z(z)\equiv \int dxdy\,\rho(\vec{r})$.
For the configuration (B), the nucleus has a large octupole deformation: the proton and the neutron numbers in the region of $z\geq 8$ fm are 6.41 and 9.50, respectively, close to $^{14}$C.
Table 1 summarizes the results of the HF + BCS
calculations for the configurations (A) and (B), in which
the deformation parameters are defined as
\begin{equation}
\beta_\lambda=\frac{4\pi}{3A} \frac{Q_{\lambda}}{\bar{R}^\lambda},
\end{equation}
with $\bar{R}=\sqrt{\frac{5\langle r^2\rangle}{3A}}$, where
$\sqrt{\langle r^2\rangle}$ is the root-mean-square matter radius
and $Q_{\lambda}=\sum_{i} r_i^{\lambda}Y_{\lambda0}(\theta_i)$ is
the mass multpole moments.

\begin{table}[hbtp]
  \caption{
The results of the Skyrme Hartree-Fock+BCS calculations for
the $^{222}$Ra nucleus.
The table summarizes the calculated values for
the quadrupole and octupole deformation parameters, $\beta_2$ and $\beta_3$,
the root-mean-square (rms) matter radius, $\sqrt{\langle r^2\rangle}$,
the rms radius of protons, $\sqrt{\langle r_p^2\rangle}$, and
the total energy,
at two different configurations (A) and (B)
shown in Fig. 1.
}
  \label{table:Ra}
  \centering
  \begin{tabular}{ccccccc}
    \hline
    \hline
    interaction  & config. & $\beta_2$  &  $\beta_3$ &$\sqrt{\langle r^2\rangle}$
&$\sqrt{\langle r_p^2\rangle}$ & $E$ \\
    &  &  &  & (fm)
& (fm) & (MeV) \\
    \hline
    SkM* & (A)  &0.229 &0.000  &5.783&5.695&$-$1697.538\\
         & (B)  &0.466 &0.553  &6.194&6.125&$-$1672.766\\
\hline
    SLy4 & (A)  &0.209 &0.000  &5.770&5.686&$-$1695.376\\
         & (B)  &0.464 &0.553  &6.196&6.127&$-$1667.459\\
    \hline
    \hline
  \end{tabular}
\end{table}

We next solve the Hill-Wheeler equation (\ref{eq:HWeq})
in the region of $0 \leq Q_3 \leq Q_t=15000$ fm$^3$ with the
mesh size of $\Delta Q_3=\frac{15000}{7}$ fm$^3$.
We have confirmed that the GCM spectrum is almost converged
with this mesh size, and moreover,
the order of magnitude for the decay rate
remains the same even if we employ
$\Delta Q_3=2000$ fm$^3$ or $\Delta Q_3=2400$ fm$^3$.
The effect of the number projection is found to be minor,
altering the decay rate only by a factor of 2 or smaller.
See Table II for the actual values of the decay rates.

Figure 4 shows the square of the collective wave function,
$|g(Q_3)|^2$,
for the lowest GCM state.
As the octupole moment $Q_3$ increases,
the absolute value of the collective wave function
decreases and
the value of $S=|g(Q_t)|^2$ at $Q_3=Q_t$
is in order of $10^{-10}$ for both the interactions.

\begin{figure}
\includegraphics[width=8.6cm]{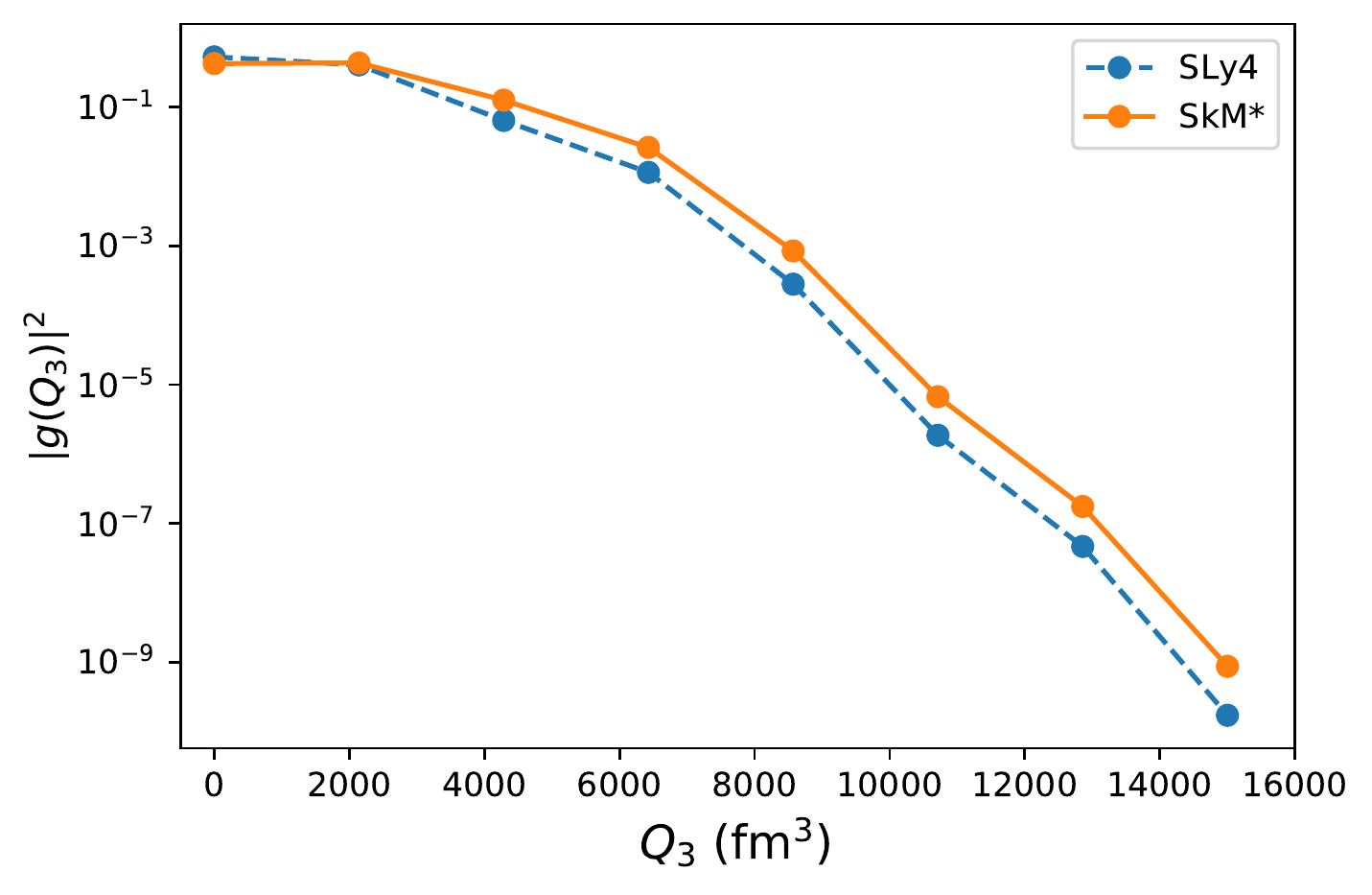}
\caption{
The square of the collective wave function
for the GCM ground state of the $^{222}$Ra nucleus as
a function of the octupole moment $Q_3$.
The solid and the dashed lines show the results with the
SkM$^*$ and SLy4 interactions, respectively.
}
\end{figure}

\begin{figure}
\includegraphics[width=8.6cm]{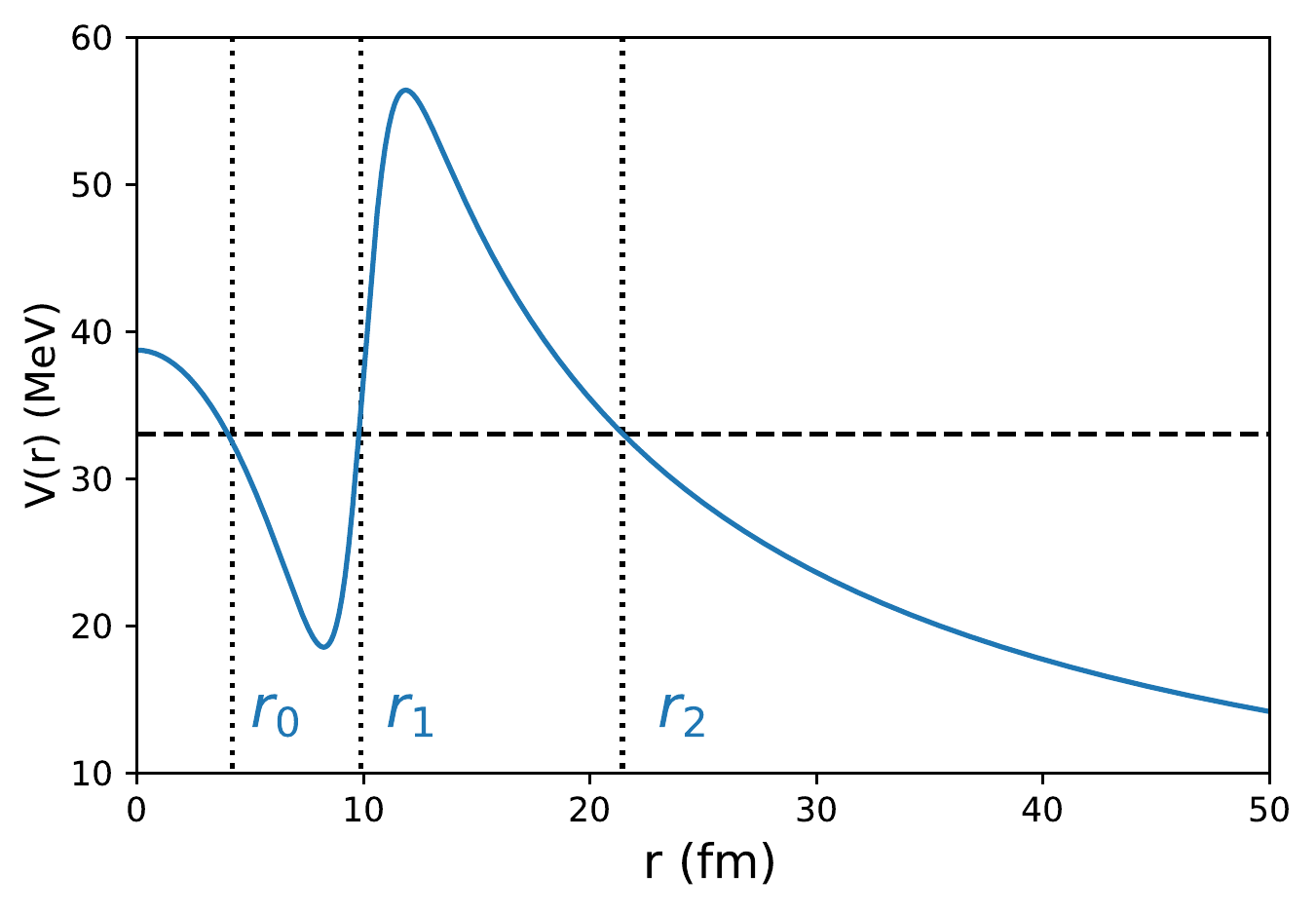}
\caption{The potential energy
between $^{14}$C and $^{208}$Pb
as a function of the relative distance $r$.
The classical turning points are denoted by $r_0,~r_1$, and $r_2$.
The dashed line denotes the $Q$-value ($Q$=33.05 MeV) for the
cluster decay of $^{222}$Ra.
}
\end{figure}

We next evaluate the assault frequency $f$ and the penetrability $P$
based on the potential model as described in Sec. II.
The potential between the two fragments is shown in Fig. 5, after
the depth parameter of the nuclear interaction is adjusted to
reproduce the resonance energy.
The resultant depth parameter is
$V_0=67.64$ MeV, whereas the radius and the diffuseness parameters are
$R_0=9.99$ fm and $a=0.63$ fm, respectively.
The dashed line corresponds to the $Q$ value of the cluster decay.
With this parameter set, the assault frequency and the penetrability 
are found to be $f=8.29\times10^{20}$ s$^{-1}$ and 
$P=4.10\times10^{-26}$, respectively.

\begin{figure}
\includegraphics[width=8.6cm]{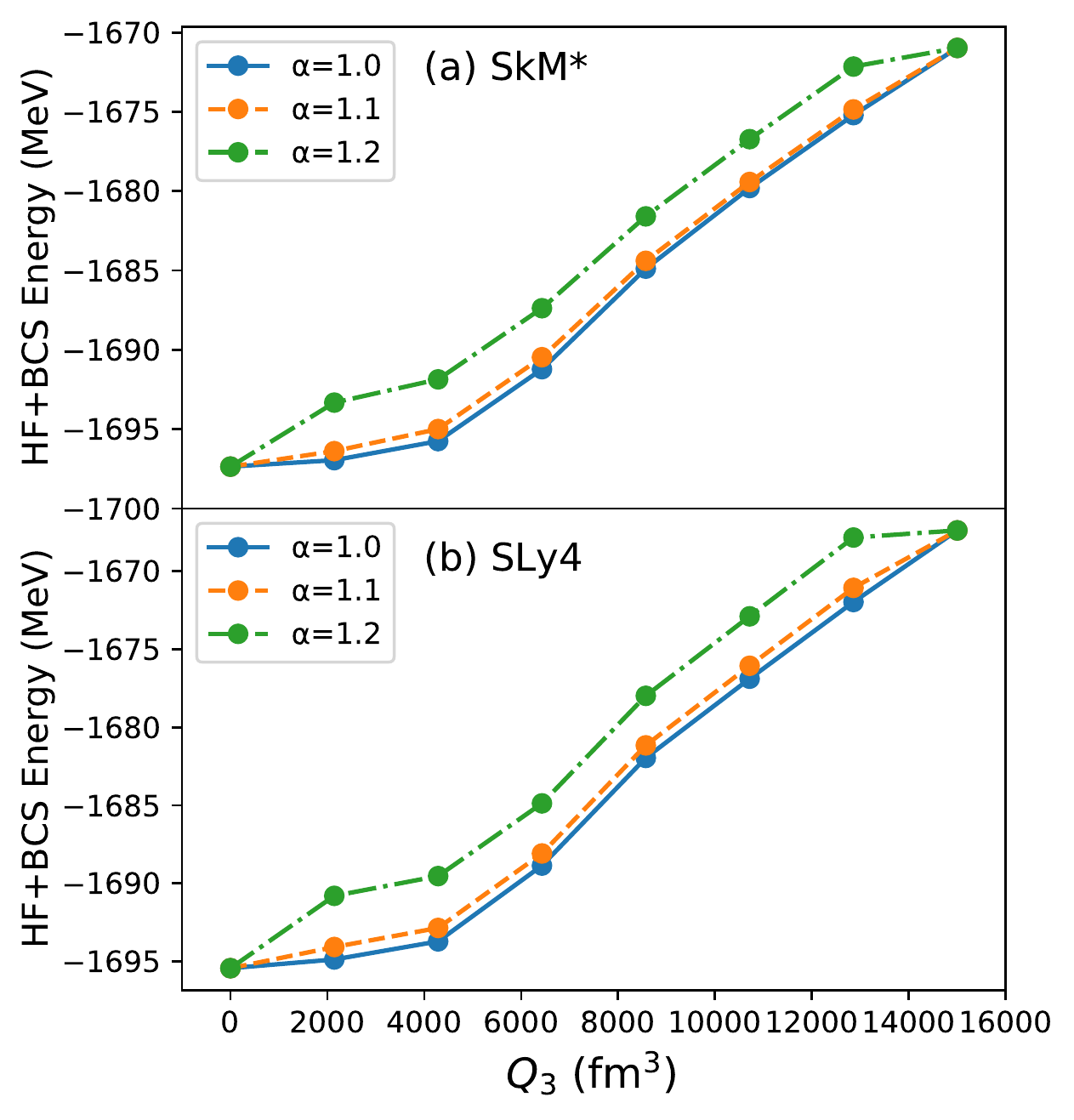}
\caption{
Similar to Fig. 1, but obtained
with the BCS wave function for the modified Hamiltonian, Eq. (\ref{eq:Hmod}).
Notice that the energy shown is defined as the expectation value of the
original Hamiltonian with $\alpha=1$.
The upper and the lower panels show the results of the SkM$^*$ and
the SLy4 interactions, respectively.
}
\end{figure}

In order to investigate the effect of dynamical pairing,
we next apply the maximum coupling approximation\cite{hagino2020b}.
Figure 6 shows the HF+BCS energy for three different values of $\alpha$ in
Eq. (\ref{eq:Hmod}).
Notice that we
fix the value of $\alpha$ to be 1 for the configurations at
$Q_3=0$ and $Q_3=Q_t$, while for the other configurations
we solve the HF+BCS equations with the modified Hamiltonian.
The expectation value of the original Hamiltonian is then computed
with the wave functions so obtained.
Since such wave functions contain excited states
components (see Eq. (\ref{eq:thouless})),
the total energy increases for $\alpha\neq 1$.
On the other hand, the off-diagonal components of the overlap kernel
tends to be increased when $\alpha$ is varied from one.
These two effects compete with each other in evaluating the
decay rates.

Figure 7(a) shows the preformation probability
$S=|g(Q_{t})|^2$ as a function of $\alpha$.
The corresponding decay rate $w$ is shown in Fig. 7(b).
Because of the competition of the two opposite effects mentioned in the
previous paragraph, a peak
structure appears in the decay rate,
at $\alpha=1.075$ and $\alpha=1.1$
for the SkM$^*$ and the SLy4 interactions, respectively,
similar to the previous studies on the
role of dynamical pairing
in spontaneous fission \cite{giuliani2014,guzman2018,hagino2020b}.
The decay rate at the peak is larger than the decay rate for the
original pairing strength (i.e., $\alpha=1$)
by a factor of 1.82 and 2.04 for the SkM$^*$ and
SLy4, respectively.
Notice that the decay rate with SkM$^*$ is larger than that with SLy4.
This is because
the energy difference between the configurations (B) and (A) is larger
with SLy4 (see Table I).

\begin{figure}
\includegraphics[width=8.6cm]{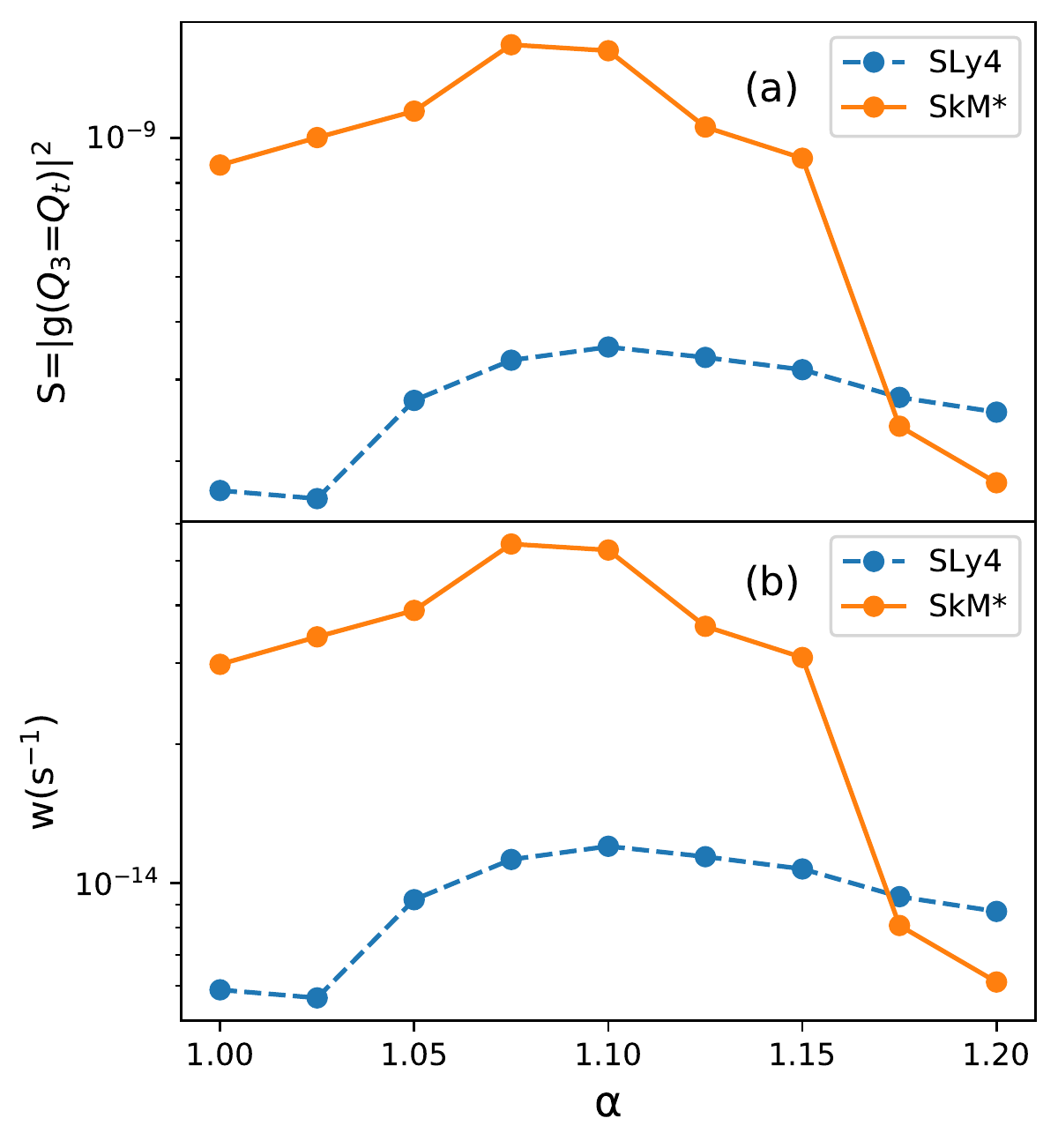}
\caption{The preformation probability (the upper panel)
and the decay rate (the lower panel)
for the cluster decay $^{222}$Ra$\rightarrow^{208}$Pb+$^{14}$C
for various values of $\alpha$ for the maximum coupling
approximation.
}
\end{figure}

The decay rates at the maxima in Fig. 7(b) are summarized in Table II
together with the experimental data.
For comparison, the calculated result without the number projection
is also listed.
The present calculations reproduce the experimental deta within two orders
of magnitude, that would be reasonable as a microscopic calculation for fission.
In the Table, we also compare our results to the
calculated result of Ref. \cite{warda2011} with the Gogny D1S interaction,
which uses the model based on the
WKB approximation for spontaneous fission with the least action path.
One can see that the degree of agreement of our results with the data
is comparable to that of the result of
Ref. \cite{warda2011}.

\begin{table}[hbtp]
\caption{
The decay rates $w$ for the cluster decay of
$^{222}$Ra $\rightarrow ^{208}$Pb+$^{14}$C
obtained with the present approach. They are compared to
the experimental data and to the theoretical calculation \cite{warda2011}
based on the WKB model with the least action fission path.
  }
  \label{table:decayRa}
  \centering
  \begin{tabular}{ll}
    \hline
    \hline
    $w~(\mathrm{s}^{-1})$ & the method \\
    \hline
     $5.43\times 10^{-14}$  & GCM (SkM*)   \\
     $8.15\times10^{-14}$  & GCM without the projection (SkM*)    \\
     $1.20\times10^{-14}$  & GCM (SLy4)    \\
     $8.73\times10^{-10}$    & the least action (Gogny D1S) \cite{warda2011} \\
     $6.7~(\pm1.8)\times10^{-12}$    & Price {\it et al.}\cite{price1985}  \\
     $5.6~(\pm2.2)\times10^{-12}$    & Hourani {\it et al.} \cite{hourani1985}\\
     $4.20~(\pm1.18)\times10^{-12}$    & Hussonnois {\it et al.} \cite{hussonnois1991}\\
    \hline
    \hline
  \end{tabular}
\end{table}

\subsection{$^{228}\mathrm{Th}$ and $^{232}\mathrm{U}$}

We next discuss the cluseter decay of $^{228}$Th $\rightarrow ^{208}$Pb+$^{20}$O
and $^{232}$U $\rightarrow ^{208}$Pb+$^{24}$Ne.
The $Q$-values of these decays are 44.72 MeV and 62.33 MeV.
The octupole moment at $Q_t$ reads
$Q_t=2.0\times 10^4$ fm$^3$ and $2.4\times 10^4$ fm$^3$ for
$^{228}$Th and $^{232}$U, respectively.
The calculated decay rates are shown in Fig. 8 as a function of $\alpha$
for the maximum coupling approximation.
To this end, we use the mesh size of
$\Delta Q=2000$ fm$^3$ to discretize the Hill-Wheeler equation.
For these nuclei, the collective wave functions at $Q_3=Q_t$ are as small as the order of
10$^{-6}$, and thus it easily suffers from numerical instabilities.
In order to avoid this,
we extrapolate the wave function in the region of
6000 $\leq Q_3 \leq 12000$ fm$^3$ down to $Q_t$.
The qualitative features of the decay rates are the same as those for
$^{222}$Ra discussed in the previous subsection.
That is, the decay rate is enhanced by several times by introducing the effect
of dynamical pairing, and the calculated decay rates
reproduce the experimental data within the same order
for $^{228}$Th and within two or three orders of magnitude for $^{232}$U.
See Table III for a summary of the calculated results.

\begin{figure}
\includegraphics[width=8.6cm]{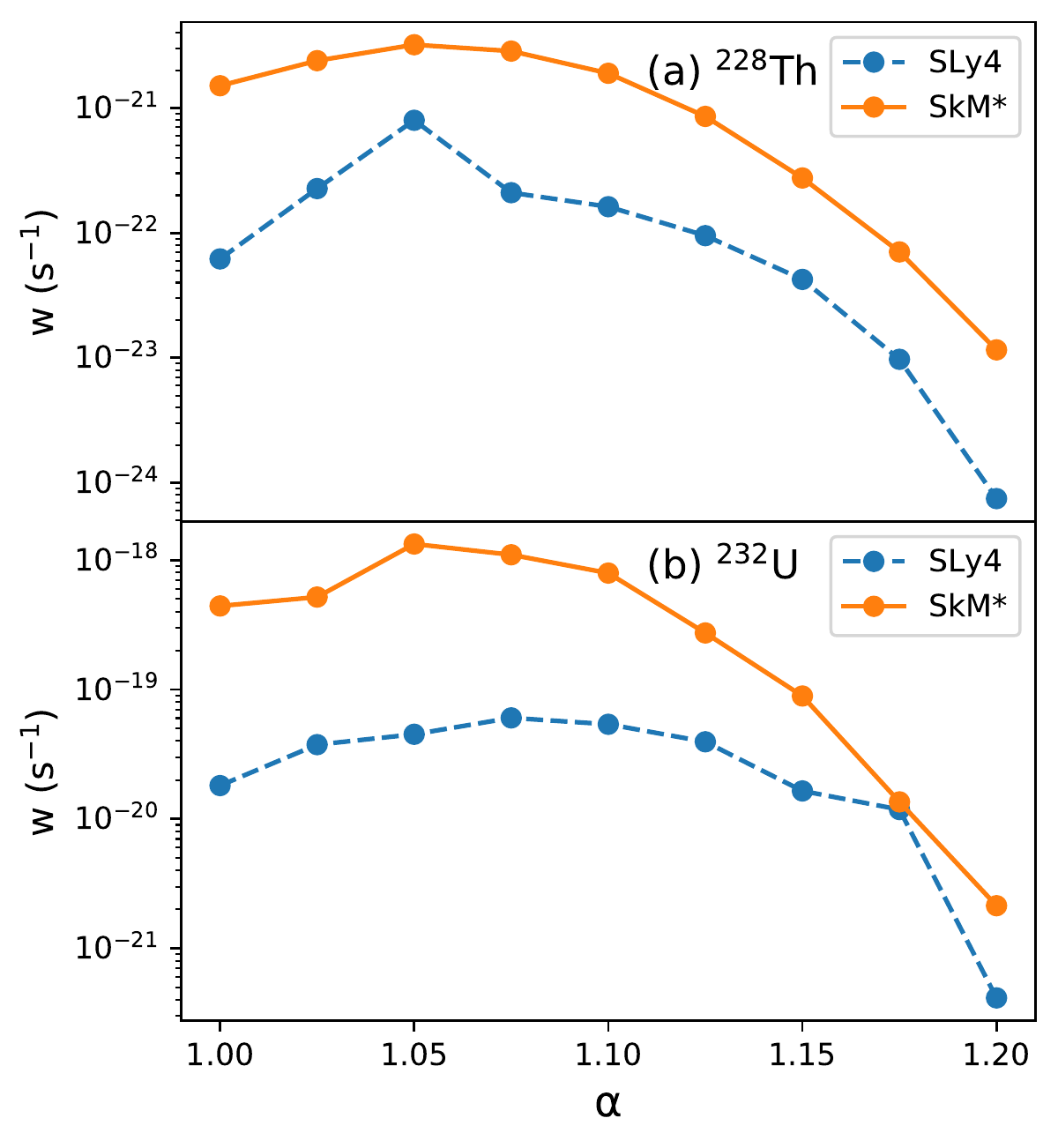}
\caption{Same as Fig. 7(b), but for
$^{228}$Th $\rightarrow ^{208}$Pb+$^{20}$O (the upper
panel) and $^{232}$U $\rightarrow ^{208}$Pb+$^{24}$Ne (the
lower panel).
  }
\end{figure}
\begin{table}[H]
  \caption{
Same as Table II, but for
$^{228}\mathrm{Th} \rightarrow ^{208}\mathrm{Pb}+^{20}\mathrm{O}$ and $^{232}\mathrm{U} \rightarrow ^{208}\mathrm{Pb}+^{24}\mathrm{Ne}$.
  }
  \centering
  \begin{tabular}{c|ll}
\hline
\hline
the nucleus &    $w~(\mathrm{s}^{-1})$& the method\\
    \hline
$^{228}$Th &    $3.20\times10^{-21}$ & GCM (SkM*) \\
    & $7.96\times10^{-22}$  &GCM (SLy4) \\
    & $2.05\times10^{-20}$    &the least action \cite{warda2011} \\
    &    &(Gogny D1S)  \\
    & $1.29~(\pm0.22)\times10^{-21}$    &Bonetti {\it et al.} \cite{bonetti1993}  \\
    \hline
$^{232}$U &     $1.32\times10^{-18}$&GCM (SkM*) \\
    & $6.03\times10^{-20}$&GCM (SLy4)  \\
    & $3.10\times10^{-24}$    &the least action \cite{warda2011} \\
    &   &(Gogny D1S)  \\
    & $6.3~(\pm1.5)\times10^{-22}$    & Barwick {\it et al.} \cite{barwick1985}\\
    & $2.72~(\pm0.23)\times10^{-21}$    & Bonetti {\it et al.} \cite{bonetti1990} \\
    & $2.83~(\pm0.22)\times10^{-21}$    & Bonetti {\it et al.}\cite{bonetti1991}\\
    \hline
    \hline
  \end{tabular}
\end{table}

\section{Summary}

Using the generator coordinate method, we have estimated
microscopically the preformation probabilities for the cluster
radioactivities of $^{222}$Ra, $^{228}$Th and $^{232}$U.
Unlike the pair hopping model, we have taken into account
the non-orthogonality of the configurations as well as non-nearest neighbor
couplings. Moreover, we have
employed the maximum coupling
approximation to take into account the effect of
dynamical pairing.
By combining with the Gamow theory, we have shown that the experimental
decay rate for $^{222}$Ra is reproduced reasonably well with this calculation and the same is true for $^{228}$Th and $^{232}$U even though there is some uncertainty derived from the fitting.
We have also shown that the dynamical pairing increases the
preformation probability by a factor of two or three.

In this paper, we have used the octupole moment as a generator coordinate.
In principle, one can also incorporate explicitly
other degrees of freedom, such as the quadrupole moment, the
hexadecapole moment, and the neck degree of freedom.
In particular, the neck has been known to play an important role
in treating the scission dynamics and thus a connection between a one-body
system to a two-body system. By taking into account the neck degree of
freedom, the preformation probability of a cluster may also be better defined.
This will be an interesting future problem, even though a numerical accuracy
of a GCM solution would be more demanding.

The cluster decay can be regarded as a phenomenon in between spontaneous
fission and $\alpha$ decays.
The method presented in this paper opens a novel and promising way
to develop a unified microscopic description for
quantum tunneling decays of nuclear many-body systems, including
$\alpha$ decays, cluster decays,
and spontaneous fission.
For spontaneous fission, the Gamow theory cannot be applied
in a straightforward manner, since the barrier penetration and
a formation of fission fragments are strongly coupled to each other.
Extending
the method presented in this work to such a problem
would also be
an interesting future work.

\begin{acknowledgments}

We thank George F. Bertsch for useful discussions and encouragements.
This work was supported in part by
JSPS KAKENHI Grant Numbers JP19K03824, JP19K03861, JP19K03872, and JP21H00120.
The numerical calculations were performed
with the computer facility
at the Yukawa Institute for Theoretical
Physics, Kyoto University.

\end{acknowledgments}

\end{document}